\documentstyle[11pt,paspconf,epsf]{article}
\begin{document}

\title{The Chemical Evolution of Galaxies at High Redshift}

\author{Linda J. Smith}
\affil{Department of Physics and Astronomy, University College London,
Gower Street, London WC1E 6BT, UK}

\author{Max Pettini and David L. King}
\affil{Royal Greenwich Observatory, Madingley Road, Cambridge, CB3 0EZ, UK}

\author{Richard W. Hunstead}
\affil{School of Physics, University of Sydney, NSW 2006, Australia}

\begin{abstract}
Observations of absorption lines in the spectra of distant QSOs
offer a new approach for tracking the evolution of normal galaxies
from early epochs to the present day. The damped Ly$\alpha$
systems are particularly suitable for measuring the properties of what 
are likely to be the progenitors of present-day luminous galaxies.
We have recently concluded a long-term survey of 30 damped 
absorbers (including eight from the literature)
aimed at measuring the metallicity and 
dust content of the universe from redshift $z=3.39$ to 0.69\,. 
The major conclusions are that 
the epoch of chemical enrichment in galaxies may have begun at
$z \simeq 2.5-3$---corresponding to a look-back time of $\sim
14$~Gyr ($H_{\rm 0} = 50$\,km\,s$^{-1}$\,Mpc$^{-1}$, 
$q_{\rm 0} = 0.01$)---and that at $z \simeq 2$ 
the typical
metallicity was 1/15 of solar. There is clear
evidence for the presence of interstellar dust at $z \approx 2$, although
several high-redshift galaxies, particularly the most metal-poor,
appear to be essentially dust-free.
We discuss the nature of the damped Ly$\alpha$ galaxies in the light of
these and other new results.
\end{abstract}



\section{Introduction}
Our current knowledge of galactic chemical evolution is limited to studies
of old Galactic stars and extragalactic H\,II regions. To probe the
chemical evolutionary history of galaxies in general we need to identify
normal galaxies and measure element abundances over a large range of 
redshifts.
The so-called `damped Ly$\alpha$ systems', a class of QSO
absorbers with neutral hydrogen column densities $N$(H\,I)$ > 2 \times
10^{20}$ \,cm$^{-2}$, are ideal targets for this
purpose. First, they are selected simply by their strong damped Ly$\alpha$
profiles in the spectra of background QSOs. Studies of damped Ly$\alpha$
systems (see Wolfe 1995 for a recent review) suggest that they are the
progenitors of present-day luminous galaxies and that they trace the
bulk of neutral material available for star formation at high
redshifts. 
Second, the large neutral hydrogen column density ensures that 
ionization corrections are
negligible in the calculation of element abundances (Viegas 1995). 
The typical spectrum of a damped Ly$\alpha$ galaxy
is similar to that produced by local interstellar gas seen in
absorption against a background star. Indeed, ultraviolet observations of 
interstellar gas have shown that with a careful choice of elements and 
transitions, it is possible to determine not only accurate metallicities but 
also the amount of dust present.

The Zn\,II doublet at 2025 and 2062\,\AA\ is a highly suitable tracer of the 
degree of metal enrichment for two reasons.
In Galactic stars the abundance of Zn tracks those of
Fe-peak elements down to very low metallicities 
(Sneden, Gratton, \& Crocker 1991), 
and in the diffuse interstellar medium  Zn is within $\approx$ 0.2
dex of solar (Sembach et al. 1995; Roth \& Blades 1995). 
An indication of the amount of
dust present can be obtained from transitions of Cr\,II occurring close
to the Zn\,II lines at 2055, 2061 and 2065\,\AA\ since in the
local interstellar
medium  $\approx$ 99\% of Cr is depleted onto grains. 
Moreover, the solar abundances of both Zn and Cr are low, ensuring that the
absorption lines are generally unsaturated and the corresponding column densities
can be determined relatively accurately. 
Thus the observation of the small wavelength interval
2025--2065\,\AA\ in the rest-frame of a damped Ly$\alpha$ galaxy provides
the abundance of iron-peak elements in the interstellar gas via the
$N$(Zn$^+$)/$N$(H$^0$) column density ratio, and an indication of the
amount of dust present from the $N$(Cr$^+$)/$N$(Zn$^+$) ratio.

\section{Observations}
Over the last six years we have obtained intermediate dispersion
spectra of a large sample of QSOs with damped Ly$\alpha$ systems
using the 4.2\,m William Herschel Telescope and the 3.9\,m
Anglo-Australian Telescope. Our first survey paper (Pettini et al. 1994) 
presented measurements for 15 damped systems; 
recently we have extended the sample with observations of 
seven additional systems, mostly at $ z > 2.5$\,. 
Together with data from the literature
(Meyer, Welty \& York 1989; Meyer \& York 1992;
Lu et al. 1995; Meyer, Lanzetta \& Wolfe 1995;
Steidel et al. 1995a;
Smette et al. 1995;
Prochaska \& Wolfe 1996),
the total data set now includes
30 damped Ly$\alpha$ systems over the redshift range
$z=0.69$--3.39, more than one third of the total number known
(Wolfe et al. 1996). 
Zn\,II absorption has been detected in 14 cases,
all but one at $z < 2.5$, and useful upper limits 
have been obtained for the remaining 16 sightlines.
Figure 1 shows all the available data on the abundance of Zn in damped 
Ly$\alpha$ systems as a function of redshift.

\begin{figure}
\epsfbox[114 370 456 541]{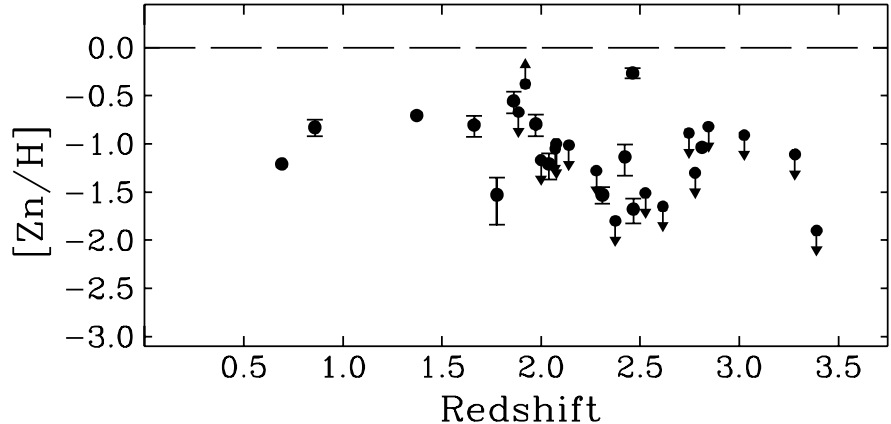}
\caption{The abundances of Zn relative to solar (on a log scale) in
the 30 damped Ly$\alpha$ galaxies in our survey plotted against redshift. Upper
limits, corresponding to the non-detection of the Zn\,II lines, are
indicated by smaller dots and downward pointing arrows.
The dashed line corresponds to the solar
abundance of Zn.}
\end{figure}

%
%
\section{Chemical Enrichment in the Early Universe}
As can be seen from Fig.~1, 
the Zn abundances are generally well below solar, indicating that most
damped Ly$\alpha$ galaxies are chemically young. At $z \approx 2$---where 
most data are available---we find an average 
$\langle{\rm [Zn/H]}\rangle = -1.2$ (1/15 solar)\footnote{This value 
is obtained by using the
Zn\,II doublet $f$-values measured by Bergeson \& Lawler (1993) and the solar
abundance of Zn from the compilation by Anders \& Grevesse (1989). For
comparison, in Pettini et al. (1994) 
we deduced $\langle{\rm [Zn/H]}\rangle = -1.0$ (1/10 solar)
based on earlier estimates of the $f$-values and of [Zn/H]$_{\odot}$\,.}.
Since at this redshift
the damped systems
dominate the mass density of neutral gas in the universe, this average
metallicity can be interpreted as the characteristic metallicity of the 
universe at a look-back time of $\approx 13$~Gyr.

Figure 1 also shows that there is a considerable range in the Zn abundance
at $z \approx 2$\,. We know that the full range spans more than
2 orders of magnitude because in several cases where only upper limits
are available for Zn,  
high resolution spectroscopy of other ions with intrinsically stronger
lines has shown that the metallicities are very low. For
example, Pettini \& Hunstead (1990) measured abundances as low as $\sim$ 1/350
solar in the $z_{\rm abs}=2.076$ system towards Q2206$-$199
([Zn/H] $ < -1.0$); similarly in the
$z_{\rm abs}=2.279$ system towards Q2348$-$147, where  
[Zn/H] $ < -1.3$, Pettini, Lipman \& Hunstead (1995) find
the abundances of Si, S and Fe to be $\approx 1/100$ of solar.
Evidently chemical
enrichment did not proceed at the same rate in different galaxies,
presumably reflecting the protracted epoch of disk formation 
(Kauffmann 1996). 

The picture at $z > 2.5$ and $z < 1.5$ is more sketchy than between these 
redshifts, reflecting the observational difficulties in detecting the 
Zn\,II absorption lines in the near-infrared and ultraviolet respectively.
Nevertheless, the available data do allow some tentative conclusions to be
reached.
In all but one case Zn is {\it un}detected at $z > 2.5$ and 
its abundance is generally lower than $-1.2$\,.
On the other hand, at $z \simeq 2 - 2.5$, some galaxies had apparently 
already reached  [Zn/H]$\geq -0.5$\,.
This suggests that the first major episodes 
of metal production in galaxies probably
occurred between $z \simeq  3$ and 2, and that in some cases  
this process may have proceeded rapidly, on a timescale of 1--2~Gyr.
This conclusion is consistent with the recent discovery of a significant
population of star-forming galaxies at $z \simeq 3$ by Steidel 
et al. (1995b, 1996).

It is perhaps 
surprising that no damped Ly$\alpha$ systems with near-solar metallicity 
have been found at $z < 1$, as we approach the epoch when the Sun formed 
(at $z \simeq 0.32$ in this cosmology). This may be 
simply due to the very small number of measurements available 
(see Fig. 1). The few damped systems imaged to date have all been found 
to be relatively underluminous galaxies (Steidel et al. 1994, 1995a).
It is also possible that at these epochs, when a significant
fraction of the gas in galaxies had been cycled through stars and the typical
content of heavy elements and dust 
had risen to values greater than $\sim 1/15$
of those found today, our spectroscopy technique
becomes significantly biased {\it against} the galaxies we are trying
to detect.
Existing compilations of QSOs with damped Ly$\alpha$ systems are all drawn 
from magnitude limited optical surveys; interstellar dust may well result in
sightlines through chemically unevolved, and therefore 
relatively unreddened, galaxies being over-represented (Pei \& Fall 1995).
While this selection effect must be operating at some level, it remains 
to be established with future observations how important it really is in 
biasing our view of the universe. 

In their study of the chemical evolution of the Galactic disk, Edvardsson
et al. (1993) concluded 
that the age-metallicity relation is relatively flat and 
shows a large scatter at all ages. If, furthermore, 
disk formation in the universe was 
not a coeval process but, as seems more likely, took place over a 
protracted epoch (Kauffmann 1996), we would not expect a tight trend in 
plots such as that shown in Fig.~1.  Nevertheless, even the broad 
characteristics of the chemical history of damped Ly$\alpha$ 
galaxies appear to be significantly different from those of the disk of 
the Milky Way. In particular, the distribution of metallicities 
at $ z \simeq 2$ resembles more closely that of stars in the halo, rather 
than the thick or thin disk (see for example Fig.~16 of Wyse \& Gilmore 
1995). On the basis of the chemical evidence we would conclude that
at $z \simeq 2$ most galaxies had not yet collapsed to 
form disks and that the damped Ly$\alpha$ systems trace an earlier stage 
in the formation of galaxies, possibly to be identified with the 
spheroidal component. It will be very interesting to 
assess whether other lines of 
evidence, particularly the kinematics 
and the morphology of high redshift galaxies
(now accessible with the Hubble Space Telescope),
support this conclusion or not.

\section{Dust in Young Galaxies}
Turning now to Cr, we find that this element is generally less abundant 
than Zn by factors of up to $>5$. 
This may reflect an intrinsic departure from solar relative abundances, 
but such an effect is seen only in the most metal-poor stars in our Galaxy, 
with [Fe/H] $< -2.5$ (McWilliam et al. 1995).
By analogy with the local interstellar medium, 
we consider it more likely that 
a fraction of the gaseous Cr 
has been incorporated in 
dust grains. Typically, 
we find that about half of the Cr is `missing' from the gas phase.
If this also applies to other grain constituents,
then the dust-to-gas ratio at the typical metallicity of 1/15 solar
is $\approx 1/30$ of that of the Milky Way.
This low value is consistent 
with the mild reddening found towards background
quasars with damped Ly$\alpha$ systems in their spectra
(Pei, Fall \& Bechtold 1991). 
If we compare our Cr/Zn ratios with those
measured in the local interstellar medium (see Fig.~7 of Pettini et al. 1994),
we find that the gas phase abundance of Cr is one order of
magnitude higher in the damped systems. This points 
to significant differences in the physical processes which determine 
the balance between gas and dust in the interstellar media of these 
high-redshift galaxies, compared to the Milky Way today.  

There is  a hint in our survey
that the Cr depletion may decrease with decreasing metallicity. 
Thus, in the lowest metallicity systems ([Zn/H]$ <-1.7$), we may expect 
to detect preferentially the Cr\,II lines since Cr is
$\sim 11$ times more abundant
than Zn in the Sun.

\begin{figure}
\epsfbox[120 299 442 462]{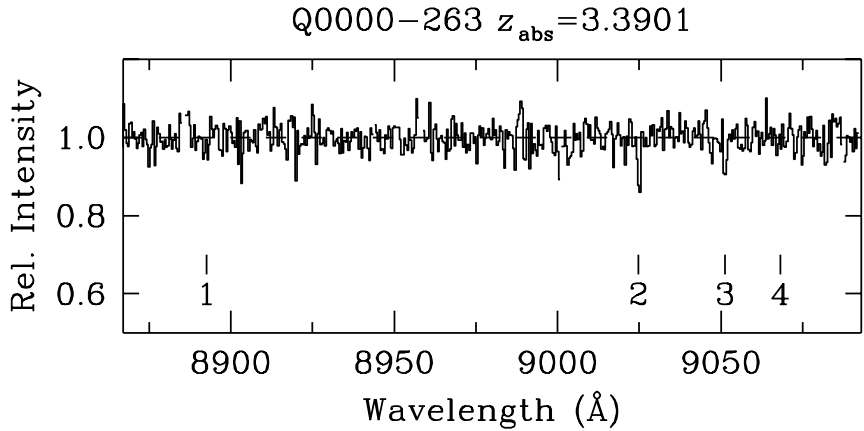}
\caption{Portion of the AAT spectrum of Q\,0000$-$263
showing  Cr\,II absorption in the
damped Ly$\alpha$ system at $z=3.3901$.
The vertical tick marks indicate the expected locations of absorption
lines of interest, whether they are detected or not.
Line 1: Zn\,II 2025.5;
line 2: Cr\,II 2055.6; line 3: Cr\,II 2061.6 $+$ Zn\,II 2062.0 (blend);
and line 4: Cr\,II 2065.5\,.
This spectrum has S/N = 30 
(note the expanded vertical scale) and 
FWHM = 1.1 \AA. }
\end{figure}

One such case is shown in Fig. 2. In our high S/N AAT
spectrum of Q0000$-$263, Zn\,II lines in the damped absorber at
the $z=3.3901$ remain undetected, implying 
an upper limit to the Zn abundance 
[Zn/H] $ \leq -1.90$ ($\leq 1/80$ solar).  However,
Cr\,II absorption is clearly present and
we deduce [Cr/H] $=-2.2 \pm 0.1$, or only 
1/150 solar. This is similar to the values reported by Molaro et al. (1996)
from high resolution spectroscopy of several other ions, suggesting that
the interstellar gas in this galaxy is essentially
dust-free. Steidel \& Hamilton (1992) have imaged the absorber
and identified it with a luminous galaxy ($ L \approx 3\,L_*$) of dimensions
10--20\,h$^{-1}$\, kpc. 
Evidently, here we have an example of a `normal'
galaxy at a very early stage of 
evolution.

In summary, it is clear that through the damped Ly$\alpha$ 
absorption systems we have a direct view of the early stages in the 
evolution of galaxies.
A substantial body of data is now being obtained on chemical enrichment, 
relative abundances of different elements, kinematics, and morphologies---and 
on how these properties evolve over cosmological
timescales.
Taken together, these new observations 
should provide a detailed picture
for comparison with current theories of galaxy
formation.\\

\acknowledgements{We are grateful to the AAT and WHT time assignment 
panels for their generous support of this demanding observational 
programme.}

\end{document}